# pykanto: a python library to accelerate research on wild bird song

Nilo Merino Recalde [1,*]

[1] Department of Biology, University of Oxford, Oxford, UK
* Corresponding author: nilo.recalde@biology.ox.ac.uk

## Abstract

Studying the vocalisations of wild animals can be a challenge due to the limitations of traditional computational methods, which often are time-consuming and lack reproducibility. Here, I present pykanto, a new software package that provides a set of tools to build, manage, and explore large sound databases. It can automatically find discrete units in animal vocalisations, perform semi-supervised labelling of individual repertoires with a new interactive web app, and feed data to deep learning models to study things like individual signatures and acoustic similarity between individuals and populations. To demonstrate its capabilities, I put the library to the test on the vocalisations of male great tits in Wytham Woods, near Oxford, UK. The results show that the identities of individual birds can be accurately determined from their songs and that the use of pykanto improves the efficiency and reproducibility of the process.

## Introduction

Collecting large amounts of acoustic data from wild bird populations has traditionally been very difficult. Due to technical limitations, studies have often been constrained to tens of individuals and tens, or at best hundreds of vocalisations. But this has changed rapidly within the last decade: compact and economic autonomous recording units, such as the AudioMoth (Hill et al. 2019), now make it possible to collect orders of magnitude more data from many more individuals at once—and to do so much more cheaply. As a direct consequence, many of the computational tools traditionally employed with bioacoustic data have quickly become obsolete: they require manual curation, segmentation, and labelling of data, which are extremely time-consuming and prone to errors.

To illustrate this point, as part of our research on a wild population of great tits (*Parus major*), we record around 50,000 songs every year, which translates to well over half a million discrete acoustic units. Any analysis that required finding, labelling, and characterising them, if done manually—as is still often the case in wild bird vocalisation research (Beecher et al. 2020; Demko and Mennill 2018; McLean and Roach 2020; Pipek et al. 2018; Youngblood and Lahti 2022)—would take a very long time to complete. This bottleneck, in turn, severely limits researchers' ability to ask questions that require large datasets to answer—such as those about social learning, vocal development, large-scale cultural diversity, and the syntactic organisation of animal vocalisations (Aplin 2019; Kollmorgen et al. 2020; Lachlan et al. 2018; Sainburg et al. 2019).

In addition to concerns over the scalability of existing data analysis pipelines, there is now a demand for tools that are freely accessible and promote transparent, reproducible research. Existing proprietary software, such as the widely used Raven Pro (up to $800, K. Lisa Yang Center for Conservation Bioacoustics 2019) and Avisoft-SASLab Pro (up to $2,835, Specht 2002) are difficult to reconcile with contemporary data science practices that rely on open-source programming languages such as R (R Core Team 2021) and Python (van Rossum 1995). There exist some excellent open-source options, such as Luscinia (Lachlan 2016), Sound Analysis Pro (Tchernichovski et al. 2000) and the more recent Koe (Fukuzawa et al. 2020). However, these were generally not designed to cope with large volumes of data, and their reliance on point-and-click graphical user interfaces limits their flexibility and hinders reproducibility.

As a response to the need for scalable and open-source tools for vocalisation data analysis and related issues, the field of bioacoustics has recently started to experiment with a new suite of methods based on deep-learning artificial neural network architectures, the same that excel at, for example, computer vision and speech recognition tasks (Stowell 2021). Segmentation and annotation pipelines based on deep neural networks have already been shown to work well in laboratory settings, where three conditions hold: i) acoustic data have a high signal-to-noise ratio, ii) there are orders of magnitude more examples per vocalisation type than there are vocalisation types, and iii) vocalisations are produced by relatively few individuals (fewer than ten to a few tens) that do so in a stereotyped manner (Coffey et al. 2019; Cohen et al. 2022; Steinfath et al. 2021). Unfortunately, none of these conditions tend to be the case in field studies, and this creates a barrier to the adoption of new methods by researchers working with natural populations.

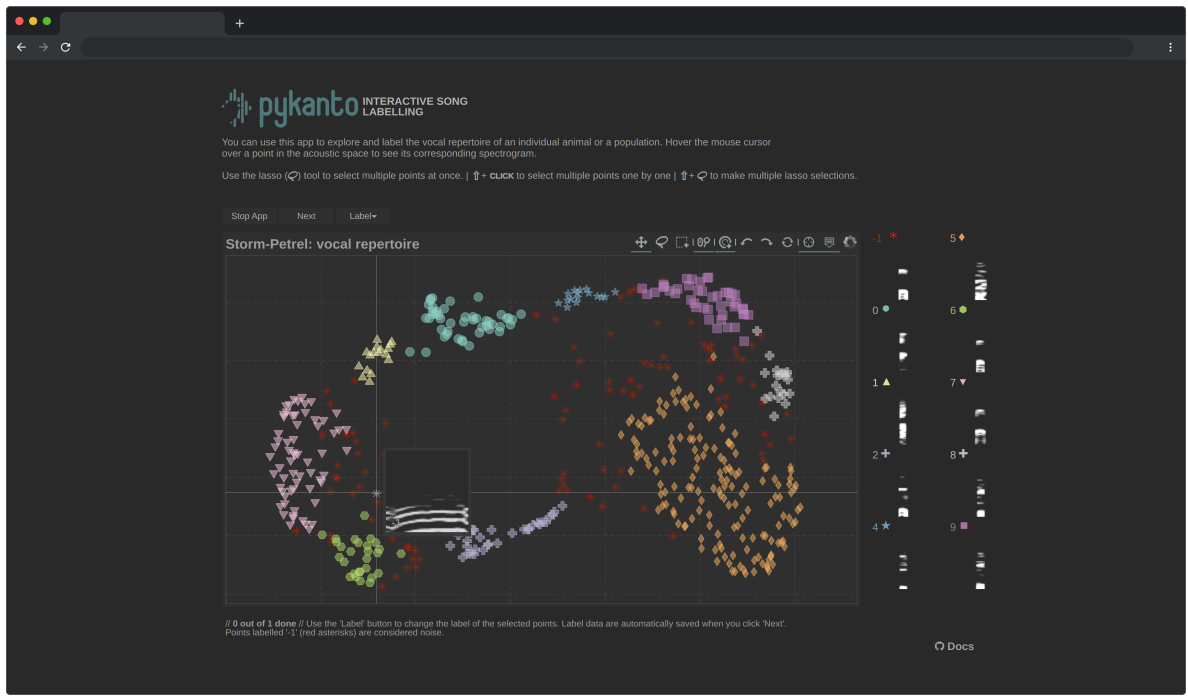

**Figure 1** Interface of the interactive web app in pykanto. This app can be used to explore datasets as well as to review and correct automatically assigned class labels in bulk.

This is the context in which I present `pykanto` (pronounced pɪˈkæntəʊ). This software library was born of three needs, which can be summarised as follows.

**First**, it needed to provide the infrastructure necessary to catalogue, explore and label large acoustic datasets collected in often suboptimal field conditions.

**Second**, it had to serve as a flexible starting point that would allow researchers to perform both traditional analyses (such as extracting hand-picked features from the vocalisations) and to use machine learning algorithms to learn low-dimensional representations of the data (Goffinet et al. 2021; Kollmorgen et al. 2020; Morfi et al. 2021; Sainburg et al. 2020), train classifiers, or detect vocalisations in unseen recordings (Cohen et al. 2022; Kahl et al. 2021; Stowell and Plumbley 2014).

**Third**, I wanted to build a tool that was free, open source, followed sustainable software practises, and geared towards computational reproducibility and transparency.

## pykanto: Implementation

`pykanto` is a software library designed to streamline the process of analysing animal vocalisations. It is programmed in Python and offers various modules to assist users in their work (see Figure 2). The central module is `pykanto.dataset` , which serves as a database for vocalizations and includes methods to visualise, segment, and label them. The `pykanto.signal` module provides tools for signal processing and creating spectrograms, while `pykanto.parameters` contains classes and functions for managing parameters. The web application `pykanto.app` allows users to explore and label large numbers of vocalizations (Figure 1) and `pykanto.plot` provides functions for plotting spectrograms. Finally, `pykanto.utils` includes parsers, I/O tools, custom typing, and general computing functions. The documentation for `pykanto` is available at nilomr.github.io/pykanto

## Dependencies

`pykanto` was written in Python 3.8 and tested in Python 3.8, 3.9 and 3.10. Its interactive web application also relies on JavaScript, HTML, and CSS. External dependencies are automatically downloaded during package installation (see the `pyproject.toml` file for a full list of dependencies).

## API and documentation

`pykanto` is a well-documented code library, making it easier to use and contribute to its development. The methods and functions in `pykanto` have clear and concise documentation, including type annotations and descriptions of their intended use. Its API (Application Programming Interface) reference, along with tutorials and practical examples, can be found in the online documentation at nilomr.github.io/pykanto.

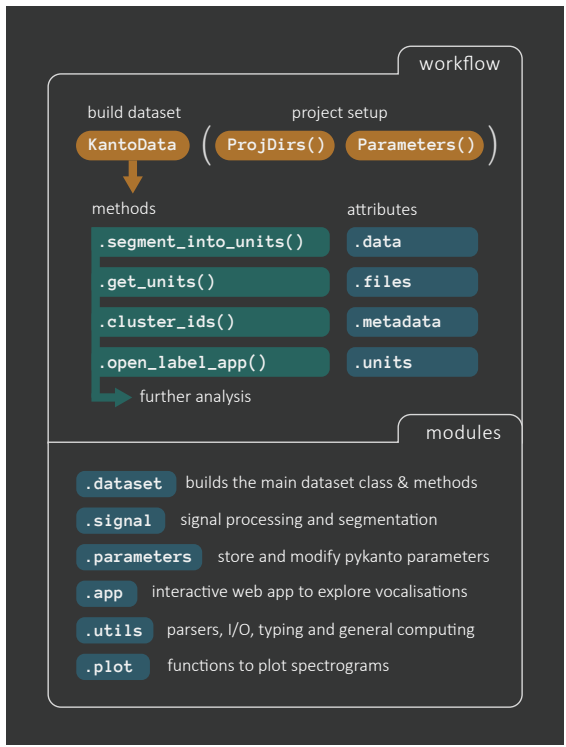

**Figure 2** `pykanto` is written around a central dataset class, `KantoData`, which provides methods to segment, visualise and label vocalisations. The library contains six modules with functions and classes to carry out common tasks in animal vocalisation analysis.

**Reproducibility and open research**

`pykanto` encourages the user to create reproducible data science projects. For example, one of its modules is dedicated to creating consistent project structures, inspired by popular utilities such as [cookiecutter](). Using the library requires writing simple scripts in Python, which allows every step of the research, from data ingestion to eventual model training and reporting, to be explicitly reproduced. The documentation includes a complete user guide with examples of best practices.

The input and output files use open data formats, and all code is available under the [MIT licence]() (a simple and very permissive licence). Where applicable, we have followed the guidelines and recommendations of the Software Sustainability Institute, a UK-based facility dedicated to research software sustainability ([software.ac.uk]()).

Many of the processes that `pykanto` carries out are computationally intensive, such as calculating spectrograms, performing operations on large arrays, and running dimensionality reduction and clustering algorithms. High-level, interpreted languages—like R or Python—are notoriously slow: where possible, we have optimised performance by both a) translating functions to optimized machine code at runtime using Numba ([Lam et al. 2015]()) and b) parallelising tasks using Ray, a state-of-the-art platform for distributed computing ([Moritz et al. 2018]()). As an example, the `segment_into_units()` function can find and segment 20.000 discrete acoustic units in approximately 16s on a desktop, 8-core machine; a dataset with over half a million (556.472) units takes 132s on a standard 48-core compute node. If `pykanto` detects a suitable GPU unit and the optional dependencies are installed, algorithms such as UMAP ([McInnes et al. 2018]()) switch to their GPU implementation, which provides a 15-100x speedup ([Nolet et al. 2021](); [Raschka et al. 2020]()). The library has a module dedicated to making it easy for users to run their scripts in a high-performance computing context (for example, a university compute cluster), and its documentation includes examples of configuration and submission scripts.

**Limitations**

This final section discusses some of the main limitations of `pykanto`. Although it will hopefully offer a flexible solution for researchers, it is also limited in important ways.

**Limitation 1**: Vocalisation unit segmentation via the very simple amplitude thresholding algorithm will not work well with species whose vocalisations vary greatly in amplitude, or with very noisy datasets. In those cases, and depending on data volume, segmentation might better be performed either manually or in a semi-automated way. For example, one could use *chipper* ([Searfoss et al. 2020]()) or train a neural network like TweetyNet, ([Cohen et al. 2022]()) on a manually annotated subset of the data.

**Limitation 2**: `pykanto` has been tested on species that produce vocalisations made up of a small or moderate number of different but distinct elements (variously referred to as notes or syllables). It will be useful for researchers working with any species, but the automatic part of the clustering process will work increasingly poorly with those that have a large number of very variable elements. This is true of any clustering method: they will fail or produce spurious results if variation in the data is continuous.

**Limitation 3**: The library does not include methods to train models intended to find analysable vocalisations in long recordings of entire soundscapes. This is a particularly challenging problem ([Priyadarshani et al. 2018]()) without a universal solution. However, `pykanto` can be used to generate and organise the training data required by these models ([Kahl et al. 2021](); [Stowell et al. 2019](); [Stowell and Plumbley 2014]()), and to work with their output annotations.

**Limitation 4**: `pykanto` is intended as a flexible solution for managing and preparing animal vocalisation data for further analysis. It provides tools that can save researchers a great deal of time while making analysis pipelines more reproducible. However, it does not implement any specific analysis or feature extraction methods, since these will vary greatly by use case.

This means that researchers using the library as part of their work will need to either have or develop familiarity with bioacoustic analysis and scripting in Python.

## Using pykanto: can individual birds be identified from their songs?

I now provide a worked example of how `pykanto` can be used to help answer real research questions vocalisations—bird song in this case:

### Introduction

Great tits are small, short-lived birds (average lifespan: 1.9 years) that sing acoustically simple yet highly diverse songs. In Wytham Woods, Oxfordshire (UK), a population of these birds has been the focus of a long-term study that is now in its 75th year. For the past three years, we have recorded the song repertoires of hundreds of individual males when they sing close to their nest before their partner begins laying. With the help of these data, we are trying to answer questions about song learning and cultural change in natural populations.

To do this we first need to know which individuals are present in the breeding population for the first time, and which were already around in previous years. However, individual survival over the winter months is low and detection by traditional means—such as mist-netting or identification in the nest—is imperfect. So we would first like to test whether individual birds can be identified based on their songs alone, and then quantify how much variation in song types occurs within and between years.

Our example dataset consists of 5293 songs from 12 males that were known (from physical recaptures) to be present in the breeding population in two different years, 2020 and 2021. Although this is a small subset of our data, it is large enough that it would still take weeks to process and analyse using traditional methods. We demonstrate the use of `pykanto` to a) organise, segment and label the dataset, and b) prepare it so that we can train a deep neural network to recognise song types. The entire process, which takes under an hour to complete, can be computationally reproduced using its [dedicated repository](). The repository includes raw data, auxiliary scripts and detailed instructions. Below is a short narrative description of the process.

### Running the analysis

**Installation** `pykanto` can be used outside a virtual environment, but this is not encouraged. Using clean environments for each project will allow you to avoid dependency issues. Once inside a new environment with Python 3.8 or above, you can install `pykanto` by simply running `pip install pykanto`, then install the package containing this example. See detailed installation and use instructions in the `.README`.

**Creating a new project and dataset** Our first step will be to define a directory structure for our project and a `ProjDirs` object to hold everything together. Then, we can test and set adequate parameters for our dataset. These include things like low- and high-cut filters, spectrogram settings, amplitude thresholding, and whether the analysis will be carried out at the song or note level. The data folder in the project already contains `.wav` audio files and their corresponding `.json` with annotations, so we can create a `KantoData` instance: this will be our database.

**Segmenting songs and using the interactive app** Then, using the `.segment_into_units()` method, we find segment onsets, offsets, unit and silence durations and add them to `KantoData.data`, the main data frame in our database object. At this point, we could already carry out most of the analyses common in the bird song literature, for example, by extracting some simple acoustic parameters from the segmented data. Instead, we want to preserve all the temporal and spectral information that is available in the spectrograms to train a more accurate classifier.

The next step is to compute and store spectrograms for each unit under examination, and then reduce their dimensionality and group them into clusters. This can be achieved by using the `.get_units()` and `.cluster_ids()` methods, which employ algorithms such as UMAP ([McInnes et al. 2018]()) and HDBSCAN ([McInnes et al. 2017]()). Afterwards, we can launch the interactive web app by calling `.open_label_app()`. Using this app, we can review the automatic labels for up to tens of thousands of vocalisations at once, splitting or combining clusters as needed. Once completed, we will have a fully annotated dataset, which can be divided into training and testing sets and exported as labelled spectrograms using `pykanto`.

**Training a convolutional neural network classifier** The distribution of song sample sizes per individual approximately follows a power law, so there is a very large amount of data for a few birds and very little for most. We need to ensure that birds for which there is a lot of data don't skew the results and that the model doesn't learn any background noise, which would also bias the results. To do this, we use a pretrained ResNet50 backbone ([He et al. 2015]()) and progressively unfreeze earlier layers—a process known as fine-tuning—while providing the model with semi-random transformations of a small, equal-sized subset of the data for each bird and song type. The repository that supports this paper contains a streamlined way of doing this using PyTorch and PyTorch Lightning, which can be easily adapted to any other dataset.

Once the model has been trained we can test how well it can classify unseen songs drawn from the held-out dataset: in this case, we can reach a very high accuracy of around 92%. As we will describe below, most of the remaining 8% is lost to confusion between the same song types sung by the same birds in different years. Finally, we feed the rest of the dataset through the model and obtain a compressed representation of each song. We can use these to calculate how similar any pair of songs are, based on those model-learnt features that are most useful to distinguish between songs.

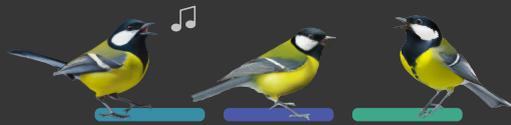
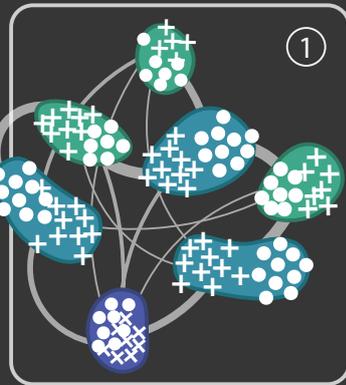
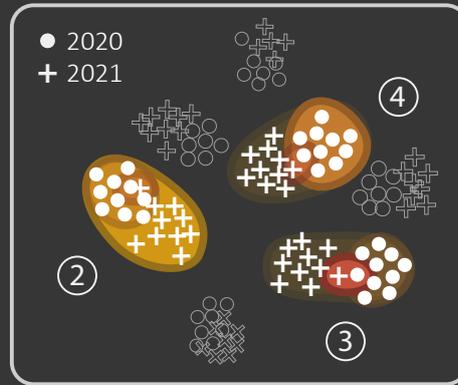
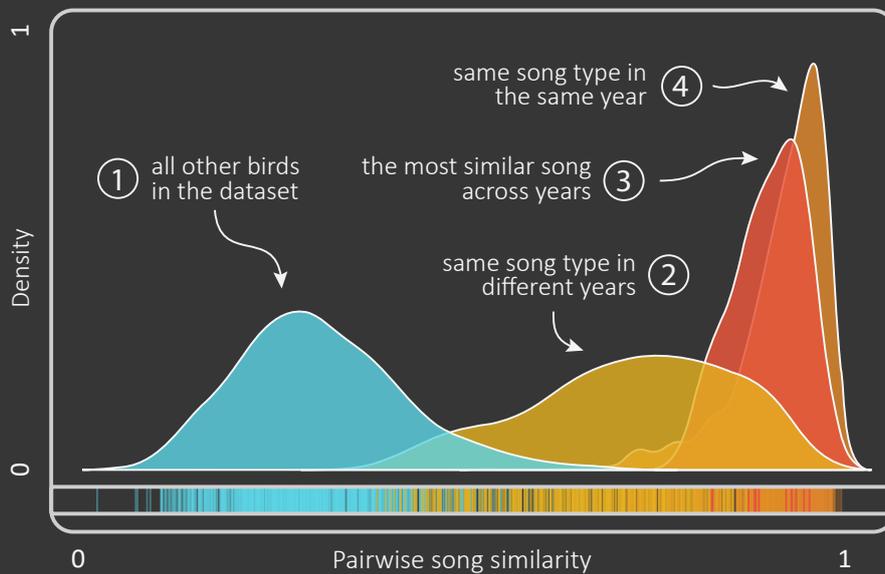

**Figure 3** (From left to right and top to bottom.) First, we calculate the pairwise similarity between the songs of all birds, which provides a baseline distribution of similarity in the population (1). Then, we compare songs from the same bird in different years (2), find the pair of songs that are most similar across years (3), and compare songs within the same year (4). The probability density estimates in the bottom panel show how pairwise similarities in (3) allow us to re-identify birds across as they do not overlap with any other birds (1).

**Results & Discussion**

Once we have calculated the similarity scores between all pairs of songs, we can partition them based on song type, bird, and year. Then, for each bird, we find which individual sang the song with the highest similarity to any within its repertoire during the following year. The baseline probability of correctly identifying the performer of the song is 2.27% (1 out of 44 song types): we successfully identified the bird in all cases. This indicates that we would have been able to re-identify the individuals even if they had not been observed or captured again.

As shown in Figure 3, the highest similarity values correspond to comparisons of the same song types within years and birds. The similarity between the same song types sung by the same bird across different years is consistently higher than that between different birds, even though some song types are shared by individuals: this means that individual vocal signatures are at least partly maintained across their lifespan.

The conclusions drawn from this analysis are limited by the small size of the dataset: including more birds would likely lead to noisier results. However, in combination with other information (such as spatial location), they might allow high-confidence identification of individuals between years without physical capture.

This example illustrates how `pykanto` can be used to help address a specific research question. The model-based vectors used to describe each song can be imported back into the KantoData database as a new column, enabling a wide range of research possibilities while maintaining a clear project structure.

**Data availability**

We distribute `pykanto` with three sample datasets that are used to run unit tests and as examples in the documentation.

**Great tit songs**: 20 songs recorded from male birds during the dawn chorus in a population in Oxford, UK. Recorded by the author and accessible at pykanto/data/great$_t$it.

**European storm-petrel purr songs**: Two males singing from burrows in the Shetland and Faroe islands. Source: XC46092 (© Dougie Preston), XC663885 (© Simon S. Christiansen). Under CC BY-NC-ND 2.5 licence.

**Bengalese finch songs**: Recordings from 2 isolated Bengalese finches. Originally published in Tachibana, Koumura and Okanoya (Tachibana et al. 2015), data can be accessed at OSF.

They can be found under `pykanto/data` when you install the package, as well as in the GitHub repository.

Additionally, the worked example in this article uses 5293 songs from male great tit songs recorded by the author between 2020 and 2021 in Wytham Woods, Oxfordshire, UK. They are available from pykanto-example/data of GitHub, along with detailed metadata.

**Code availability**

The latest version of `pykanto` is available from pip ( `pip install pykanto` ) and its source repository (github.com/pykanto). See the repository for detailed installation instructions.

`pykanto` and the example in this article rely on the following open-source scientific libraries or tools: numpy (Harris et al. 2020), scipy (Virtanen et al. 2020), pandas (The pandas development team 2023), numba (Lam et al. 2015), pytorch (Paszke et al. 2019), torchvision (TorchVision maintainers and contributors 2016), pytorch lightning (Falcon and The PyTorch Lightning team 2019), tqdm (da Costa-Luis 2019), ray (Moritz et al. 2018), soundfile (Bechtold and Geier 2022), umap (McInnes et al. 2018), joblib (Joblib Development Team 2020), hdbscan (McInnes et al. 2017), seaborn (Waskom 2021), scikit-image (van der Walt et al. 2014), librosa (McFee et al. 2015), bokeh (Bokeh Development Team 2018), ujson (van Kemenade et al. 2023), psutil (Rodola 2023), attrs (Schlawack 2019).

**Acknowledgements**

I thank Ben Sheldon and the Sheldon lab for their support and patience. Ben Sheldon, Carys Jones and Andrea Estandia provided useful comments on a draft of this manuscript. Carys Jones and Antoine Vansse tried early versions of the interactive app in `pykanto` and provided valuable feedback.

Some of the methods in `pykanto` are directly inspired by or adapted from Sainburg et al. (2020). I have indicated where this is the case in the relevant method's docstring. The dereverberation function is based on code by Robert Lachlan that is part of Luscinia (Lachlan 2016), a software for bioacoustic archiving, measurement and analysis. Please consider citing these two publications if you use `pykanto` on your own projects.

I have learnt a great deal about packaging and developing in Python by browsing the structure of existing open source projects, particularly some by David Nicholson (@NickleDave). I only became aware of VocalPy, a project that aims to "*develop an ecosystem* *of interoperable packages*" for "*computational vocal communication and learning research*" when I had already written most of `pykanto`, but eventually, I would like to make it compatible with it: standardisation is direly needed in the field and I don't want to contribute to the chaos.

This work was supported by a Clarendon-Mary Frances Wagley Graduate Scholarship and an EGI scholarship to NMR, and made use of the University of Oxford Advanced Research Computing facility (Richards 2015).

## Conflict of interest

The author declares no conflict of interest.

## Author's contributions

NMR wrote the software library and its documentation, collected the data, conducted the analyses, and wrote the manuscript.